\documentclass[reprint,aps,prd,groupedaddress,nofootinbib]{revtex4-1}

\usepackage{dcolumn}
\usepackage{amsmath}
\usepackage{amssymb}
\usepackage{bm}
\usepackage{graphicx}
\usepackage{subfig}	
\usepackage{multirow}

\begin{document}

\title{Characterization of ion-beam-sputtered AlF$_3$ thin films \\for gravitational-wave interferometers}

\author{M.~Bischi$^{1,2}$}
    \email[]{m.bischi1@campus.uniurb.it}
\author{A.~Amato$^{4}$}
\author{M.~Bazzan$^{6}$}
\author{G.~Cagnoli$^{4}$}
\author{M.~Canepa$^{7,8}$}
\author{G.~Favaro$^{6}$}
\author{D. Forest$^{3}$}
\author{P.~Gobbi$^{5}$}
\author{M. Granata$^{3}$}
	\email[]{m.granata@lma.in2p3.fr}
\author{G.~M.~Guidi$^{1,2}$}
\author{G.~Maggioni$^{6}$}
\author{F.~Martelli$^{1,2}$}
\author{M.~Menotta$^{5}$}
\author{M.~Montani$^{1,2}$}
\author{F.~Piergiovanni$^{1,2}$}
\author{L.~Valentini$^{5}$}

\affiliation{$^{1}$Universit\`{a} degli Studi di Urbino "Carlo Bo", Dipartimento di scienze pure e applicate, I-61029 Urbino, Italy}
\affiliation{$^{2}$INFN, Sezione di Firenze, I-50019 Sesto Fiorentino, Firenze, Italy}

\affiliation{$^{3}$Laboratoire des Mat\'{e}riaux Avanc\'{e}s - IP2I, CNRS, Universit\'{e} de Lyon, Universit\'{e} Claude Bernard Lyon 1, F-69622 Villeurbanne, France}

\affiliation{$^{4}$Universit\'{e} de Lyon, Universit\'{e} Claude Bernard Lyon 1, CNRS, Institut Lumi\`{e}re Mati\`{e}re, F-69622 Villeurbanne, France} 

\affiliation {$^{5}$Università degli Studi di Urbino "Carlo Bo", Dipartimento di scienze biomolecolari, I-61029 Urbino, Italy}

\affiliation {$^{6}$Università di Padova, I-35131 Padova, Italy}

\affiliation{$^{7}$OPTMATLAB, Dipartimento di Fisica, Universit\`{a} di Genova, Via Dodecaneso 33, 16146 Genova, Italy}

\affiliation{$^{8}$INFN, Sezione di Genova, Via Dodecaneso 33, 16146 Genova, Italy}

\begin{abstract}
Thermal noise in amorphous coatings is a limitation for a wide range of precision experiments such as gravitational-wave detectors (GWDs).
Mirrors for GWDs are composed of multiple thin layers of dielectric materials deposited on a substrate: the stack is made of layers with a high refractive index interleaved with layers of a low refractive index. The goal is to obtain high reflectivity and low thermal noise. 
In this paper we report on the optical and mechanical properties of ion-beam-sputtered aluminium fluoride (AlF$_3$) thin films which have one of the lowest refractive index among the known coating materials and we discuss their application in current and future GWDs.
\end{abstract}

\pacs{PACS}


\maketitle

\section{Introduction}
Brownian thermal noise in coatings (CTN) is a limiting noise source for precision experiments based on optical and quantum transducers.
For example the sensitivity of interferometric gravitational-wave detectors (GWDs) is limited in the frequency range around 100 Hz by CTN~\cite{Adhikari14, Saulson90, Levin98}. GWDs are Michelson interferometers with mirrors placed at the ends of two Fabry-P\'erot cavities, in order to measure the relative motion of the two arms. 

Mirrors for GWDs are Bragg reflectors composed by a stack of thin layers of two alternated dielectric materials (with high and low refractive index) deposited on a substrate. 
It is crucial to find coatings with low optical loss (absorption, scatter) and low thermal noise in order to improve the sensitivity of GWDs.

The power spectral density of thermally induced surface fluctuations is determined by the rate of energy dissipation in each coating material, as stated by the fluctuation-dissipation theorem~\cite{Callen52}: the higher the energy loss, the higher the CTN.

As measured with a laser beam, the CTN power spectral density $S_{\textrm{CTN}}$ can be written as \cite{Harry02}
\begin{equation}\label{eqn.S}
S_{\textrm{CTN}} \propto \frac{k_B T}{2\pi f} \frac{t_c}{w^2} \varphi_c(f, T)\ 
\end{equation}
where $k_B$ is the Boltzmann constant, $f$ is the frequency, $T$ is the temperature, $t_c$ is the coating thickness, $w$ is the laser beam radius (where intensity drops by 1/e$^2$), and $\varphi_c(f, T)$ is the loss angle associated with energy dissipation in the coating and corresponds to its  internal mechanical friction.

Thus $S_{\textrm{CTN}}$ can be reduced by increasing the beam radius $\omega$, decreasing the temperature $T$, or by choosing coating materials which minimize $t_c\varphi_c$; furthermore, the lowest CTN is expected to occur when the Young modulus of the coting ($Y_c$) matches the elastic modulus of the substrate~\cite{Harry02}. 

The reflectivity of a Bragg mirror depends on the number of low/high index layer pairs and on the refractive index contrast
 $c = n_{\textrm{\tiny{H}}}/n_{\textrm{\tiny{L}}}$, where $n_{\textrm{\tiny{H}}}$ and $n_{\textrm{\tiny{L}}}$ are the high and low refractive indices, respectively. 
 Using materials with higher $c$ allows us to use a lower number of layers and hence a lower thickness of the entire stack $t_c$.

Up to now, the high-reflectivity (HR) coatings of the Advanced LIGO \cite{aLIGO}, Advanced Virgo~\cite{AdVirgo} and KAGRA~\cite{KAGRA} GWDs are thickness-optimized stacks~\cite{Villar10} of layers of tantalum pentoxide (Ta$_2$O$_5$) doped with a relevant fraction of titanium dioxide (TiO$_2$), also known as {\it titania-doped tantala} as high index material and silicon dioxide (SiO$_2$), {\it silica}, as low index material~\cite{Granata20}. Those HR coatings are deposited by the Laboratoire des Mat\'{e}riaux Avanc\'{e}s (LMA) \cite{Degallaix19,Pinard17} using the ion-beam-sputtering (IBS) technique. Despite the excellent mechanical and optical properties of the current HR coatings \cite{Degallaix19,Granata20}, CTN remains a severe limitation for further sensitivity enhancements in GWDs. In the last two decades, a lot of  effort has been committed to find an alternative high-index material \cite{Granata20review,Vajente21}, that is the most dissipative one at room temperature, but for present and future cryogenic GWDs like KAGRA~\cite{KAGRA}, Einstein Telescope \cite{ET}, and Cosmic Explorer~\cite{Abbott17} there is a need to find alternative low-CTN materials: in fact, the currently available data on IBS silica shows a dramatic increase of dissipation below 30~K~\cite{Martin14,Granata15}, so silica coating should be replaced by another low refractive index material featuring low mechanical and optical losses.

Because of their low refractive index \cite{Kolbe92, Kolbe93, Yoshida06,Ode14} and potentially low mechanical loss at low temperature \cite{Schwarz11}, fluoride coatings could be a valid option for use in cryogenic GWDs. So far, however, fluoride coatings have never been considered for implementation in gravitational-wave detectors, so that a specifically oriented characterization of their properties is needed. The aim of this paper is to provide optical and mechanical properties of IBS aluminium fluoride (AlF$_3$) coatings as replacement of the low-index silica layers in the HR coatings of current and future GWDs. As a first step, we report the (AlF$_3$) properties measured at ambient temperature and we compare them with those of low-index silica films.

We took special care to characterize the impact of the post-deposition thermal treatment on coating properties, in particular on the $\varphi_c$ \cite{Granata20}. Annealing treatment are performed inside a tube furnace in argon-saturated atmosphere in order to avoid surface oxidation.

\section{Methods}

\subsection{Samples}
A thin layer of AlF$_3$ has been deposited by ion beam sputtering (IBS) at the Laser Zentrum Hannover \footnote{www.lzh.de} on both sides of fused silica substrates ($\varnothing$ 50 mm, $t = 1$ mm) for mechanical characterization and on one side of silicon substrates ($\varnothing$ 75 mm) for ion beam analysis, x-ray diffraction measurements and optical characterization. The coating is deposited on both sides of fused silica substrates in order to have balanced stresses and avoid deformations of the disk that could produce a variation in resonant frequencies and affect the evaluation of the frequency-dependent dilution factor.

Prior to deposition, fused-silica disks have been annealed in air at 900 $^{\circ}$C for 10 hours to release the internal stress due to manufacturing. Their masses and mechanical losses were measured so that the coating losses were calculated as explained in section \ref{SEC_mechProp}. 

Prior to deposition, the base pressure inside the coater vacuum chamber was $5 \cdot 10^{-6}$ mbar. The total pressure during the coating process was of the order of $10^{-4}$ mbar, with noble gases and gases containing fluorine injected into the chamber. Energy and current of the sputtering ions	were of the order of 1 keV and 0.1 A, respectively.

\subsection{Structure and chemical composition}
In order to determine the chemical composition of the coating samples after deposition, as well as its change after annealing treatments, we performed a series of Rutherford back-scattering spectrometry (RBS) measurements. Those measurements were done in the Accelerator AN2000 at the INFN-LNL using an $^4\text{He}^+$ ion beam at 2 MeV in normal incidence, with a scattering angle of 160$^{\circ}$.

Moreover, a series of grazing-incidence X-ray diffraction (GI-XRD) measurements were performed on coating samples after the deposition and after different annealing steps, in order to determine the coating microscopic structure and its evolution with thermal treatments. GI-XRD measurements were done with a Philips MRD diffractometer that is equipped with a Cu tube operated at 40 kV and 40 mA (collimated and partially monochromatized to the Cu K-$\alpha_1$ line using a parabolic multilayer mirror) and with a detector provided with a parallel plate collimator.

\subsection{Optical properties}
We used two J. A. Woollam spectroscopic ellipsometers to measure the coating optical properties, covering complementary spectral regions from ultraviolet to infrared: a VASE for the 0.73-6.53 ev photon energy range (corresponding to a 190-1700 nm wavelength range) and a M-2000 for the 0.74-5.06 ev range (245-1680 nm). The coated wafers were measured in reflection, their complex reflectance ratio was characterized by measuring its amplitude component $\Psi$ and phase difference $\Delta$ \cite{Fujiwara07}. To maximize the response of the instruments, the ($\Psi$, $\Delta$) spectra were acquired at different incidence angles ($\theta$ = 50$^\circ$, 55$^\circ$, 60$^\circ$) close to the coating Brewster angle. Coating refractive index and thickness were derived by fitting the spectra with realistic optical models \cite{Fujiwara07}. The optical response of the bare wafers had been characterized with prior dedicated measurements. Further details about our ellipsometric analysis are available elsewhere \cite{Amato19}.

We used photo-thermal deflection \cite{Boccara80} to measure the coating optical absorption at $\lambda=1064$ nm with an accuracy of less than 1 part per million (ppm).

\subsection{Mechanical properties}
\label{SEC_mechProp}
Two silica disks (A and B) were used for the characterization of the coating mechanical properties. 

Before and after each treatment (coating deposition, annealing runs) we measured the mass of the disks with an analytical balance and we observed the surface with an optical microscope (Olympus IX51 equipped with e ToupCam camera) to check if the coating surface was deteriorated by each treatment. Their properties are summarized in Table~\ref{TABLEproperties}.
\begingroup
\squeezetable
\begin{table}
\caption{Measured properties of silica disks for the characterization of mechanical losses.}
\label{TABLEproperties}
\begin{ruledtabular}
\begin{tabular}{lcc}
	& sample A & sample B \\ \hline
	diameter [mm] &  49.94 $\pm$ 0.02 & 49.79 $\pm$ 0.02\\
	thickness [mm] & 1.0 $\pm$ 0.1 & 1.0 $\pm$ 0.1\\
	mass before coating [mg] &  4647.3 $\pm$ 0.1 & 4612.3 $\pm$ 0.1\\
	mass after coating [mg] &  4649.1 $\pm$ 0.1 & 4613.9 $\pm$ 0.1\\
	coating thickness on each side [nm] &  211 $\pm$ 6 & 211 $\pm$ 6\\
	\colrule
\end{tabular}
\end{ruledtabular}
\end{table}
\endgroup
	
We used the ring-down method~\cite{Nowick72} to measure the frequency $f$ and ring-down time $\tau$ of several vibrational modes of each disk, before and after the coating deposition (and after each annealing treatment) in order to calculate the coating loss angle
\begin{equation}
\label{EQcoatLoss}
\varphi_c = \frac{\varphi + (D-1)\varphi_0}{D} \ , 
\end{equation}
where $\varphi = (\pi f \tau)^{-1}$ is the measured loss angle of the coated disk, $\varphi_0 = (\pi f_0 \tau_0)^{-1}$ is the measured loss angle of the bare substrate. $D$ is the frequency-dependent measured \textit{dilution factor}~\cite{Li14}:
\begin{equation}
\label{EQdilFact}
D = 1 -  \left( \frac{f_0}{f} \right)^2 \frac{m_0}{m} \ 
\end{equation}
where $m_0$, $m$ are the disk masses as measured before and after the coating deposition, respectively.
	
In order to minimize coating mechanical loss $\varphi_c$ and optical absorption $\alpha$, we tested different plateau temperatures $T_a$ and duration $\Delta t_a$. During the annealing, samples were hold in a quartz box inside the oven and maintained in a controlled argon atmosphere during the entire process. In between measurements, the samples were stored under primary vacuum ($10^{-2} - 10^{-1}$ mbar) to avoid oxidation from air exposure.
	
Disks were first measured at the Laboratoire des Mat\'{e}riaux Avanc\'{e}s (LMA) before and after coating deposition, then measured, annealed and measured again at Universit\`{a} degli Studi di Urbino Carlo Bo (UniUrb). Since the temperature in the two laboratories is different, a correction of measured mode frequencies is necessary for a correct evaluation of the frequency-dependent dilution factor, as explained in \cite{granata2021optical}.

We measured resonant modes from $\sim$2.5 to $\sim$50 kHz for each sample, in a frequency band which partially overlaps with the detection band of ground-based GWDs ($10 - 10^4$ Hz). For these ringdown measurements we used two clamp-free in-vacuum Gentle Nodal Suspension (GeNS) systems \cite{Cesarini09}. This kind of facility is currently the preferred solution of the Virgo and LIGO Collaborations for performing internal friction measurements because avoids systematic damping from suspension, clamp or residual gas pressure \cite{Granata20,Vajente17}.

The coating Young modulus $Y_c$ and Poisson ratio $\nu_c$ were estimated by fitting finite-element simulations to measured dilution factors via least-squares numerical regression \cite{Granata20}. Further details about our GeNS systems, finite-element simulations and data analysis are available elsewhere~\cite{Granata20,Granata16}.
\begin{figure}[htp]
 		\centering
 		\includegraphics[width=0.4\textwidth]{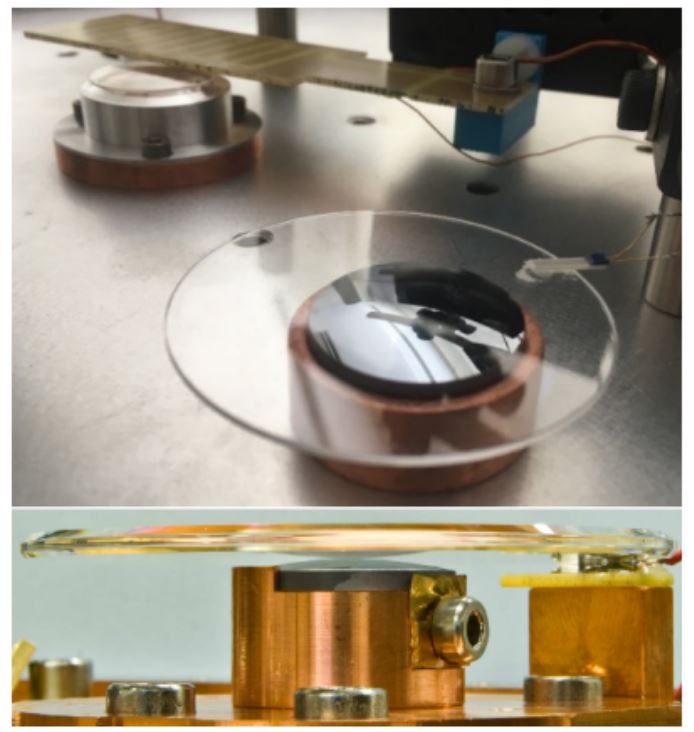}
 		\caption{GeNS systems used at the University of Urbino (top) and at LMA (bottom) to perform internal friction measurements. The sample is suspended in equilibrium on a sapphire or silicon hemisphere touching on one of the nodal points of vibration.}
 		\label{fig:GeNS}
\end{figure}

\section{Results}

\subsection{Structure and chemical composition}
\begin{figure}
 	\centering
 	\includegraphics[width=0.4\textwidth]{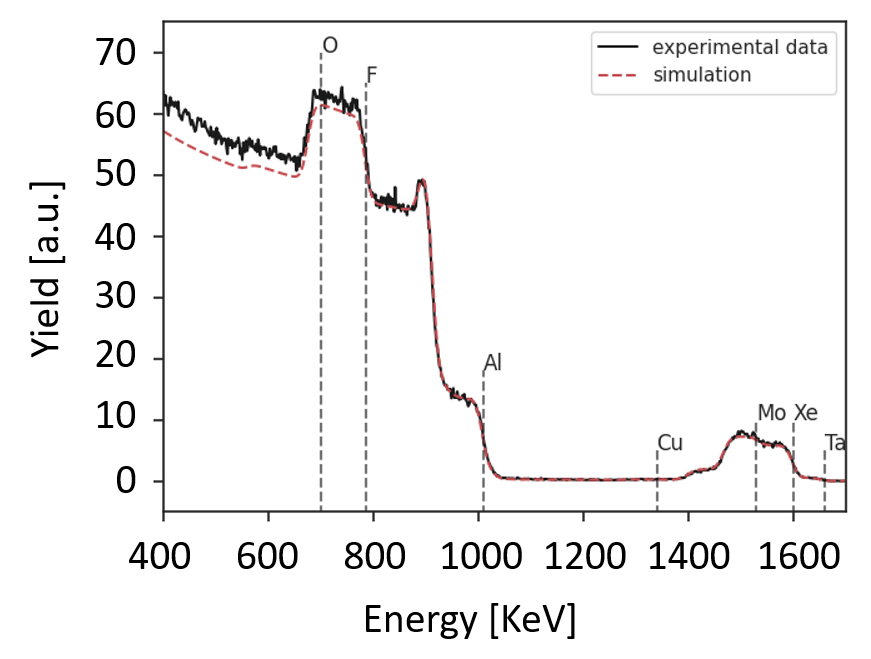}
 	\caption{Example of an RBS measurement (black line) and best fit (red line) obtained with SRIM database for stopping powers \cite{SRIM-2003}. This graph is referred to the as-deposited IBS AlF$_3$ coating.}
 	\label{FIG_RBS}
\end{figure}

RBS results of the coating samples are listed in Table II. Relative atomic concentrations and density $\rho$ are deduced from simulations of the experimental spectra based on stopping powers drawn from SRIM database \cite{SRIM-2003}. Figure~\ref{FIG_RBS} shows an example of a RBS measurement, referred to the as-deposited IBS AlF$_3$ coating.
RBS analysis of the as-deposited film shows that the Al/F atomic ratio is slightly higher than the stoichiometric one (0.4 instead of 0.33): the missing F is partially compensated by O incorporation (8\%). Moreover, the sample is contaminated by the Xe sputtering gas and by Mo from the sputtering source grids and contains traces of Cu and Ta. Film density has been calculated by dividing the areal atomic density (obtained from RBS) by the layer thickness measured via spectroscopic ellipsometry, to find a value of $\rho = 2.7$ g/cm$^3$.
\begingroup
\squeezetable
\begin{table*}[htp]
\label{TABLE_RBS}
\caption{Relative atomic concentrations (\%), Al/F atomic ratio and density $\rho$ of IBS AlF$_3$ coatings before and after thermal treatments, obtained from SRIM database for stopping powers \cite{SRIM-2003}  of the RBS spectrum. The density was calculated by assuming a layer thickness of 211 nm for all samples (measured on as-deposited samples using a spectroscopic ellipsometry).}
	\begin{ruledtabular}
		\begin{tabular}{lccccccccc}
			& Al & F & O & Xe & Mo & Cu & Ta & Al/F & $\rho$[g/cm$^3$]\\ \hline
			as deposited & 25.7 & 65 & 8 & 0.78 & 0.42 & 0.05 & 0.039 & 0.40 & 2.7\\
			200 $^{\circ}$C & 27.5 & 63 & 8 & 0.82 & 0.49 & 0.11 & 0.041 & 0.43 & 2.8\\ 
			300 $^{\circ}$C & 28.8 & 58 & 12 & 0.7 & 0.44 & 0.06 & 0.038 & 0.50 & 2.7\\
			400 $^{\circ}$C & 29.8 & 60 & 9 & 0.6 & 0.42 & 0.12  & 0.036 & 0.50 & 2.3\\ 
			500 $^{\circ}$C & 33.5 & 45 & 20 & 0.95 & 0.46 & 0.08 & 0.039 & 0.72 & 1.8\\
		\end{tabular}
	\end{ruledtabular}
\end{table*}
\endgroup

RBS analysis of the annealed samples highlights the effect of annealing treatments on the film composition. The increase of the Al/F atomic ratio, which is small at the lower temperatures, becomes particularly pronounced in the sample annealed at 500 $^{\circ}$C and is accompanied by a release of fluorine ($\sim$ 40\%) and by an increase of the incorporated oxygen (+100\%).
Film density is also affected by the annealing treatments and decreases to 1.8 g/cm$^3$ in the sample annealed at 500 $^{\circ}$C.
Concerning the other elements present in the film, annealing treatments do not seem to influence their atomic concentration, even at the highest temperature.

\begin{figure}
 	\centering
 	\includegraphics[width=0.45\textwidth]{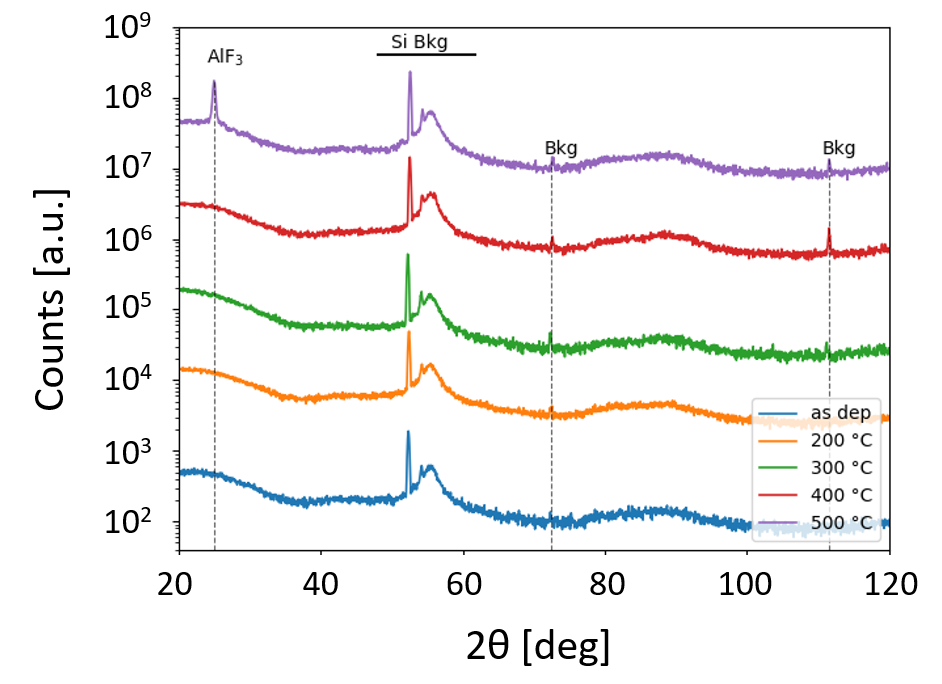}
 	\caption{GIXRD diffractograms of IBS AlF$_3$ coatings on silicon substrates, acquired before and after annealing treatments at different temperatures. Peaks between 50$^{\circ}$ and 60$^{\circ}$ are due to the background signal of the silicon substrate. The sample treated at 500 $^{\circ}$C shows an additional peak at $2\theta = 25^{\circ}$, compatible with AlF$_3$ crystalline spectrum \cite{Villars2016}.}
 	\label{FIG_GIXRD}
\end{figure}

The GI-XRD diffractograms of the coating samples are reported in Figure~\ref{FIG_GIXRD}. Peaks between 50$^{\circ}$ and 60$^{\circ}$ are due to the background signal of the crystalline silicon substrate and are visible in all the spectra.The coating material remained amorphous with the exception of the sample treated at 500 $^{\circ}$C that displays a peak at $2\theta = 25^{\circ}$, compatible with AlF$_3$ crystalline spectrum that has a main peak at 25.3$^{\circ}$ \cite{Villars2016}. The absence of other peaks is probably due to the fact that other peaks of AlF$_3$ are much less intense.
There are some peaks around 70$^{\circ}$ and 110$^{\circ}$ whose origin is not known. Their behaviour seems not to be related to annealing treatments, so they are associated to a background signal.

\subsection{Optical properties}
\begin{figure*}
 	\centering
 	\includegraphics[width=0.75\textwidth]{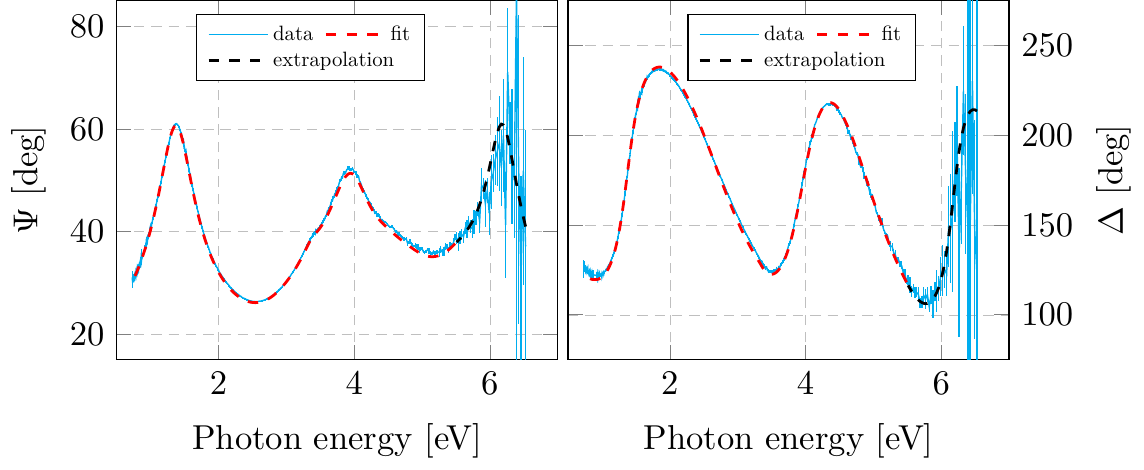}
 	\caption{Measured ellipsometric spectra of as-deposited IBS AlF$_3$ thin films, acquired at an incidence angle $\theta$ = 60$^\circ$.}
 	\label{FIG_spectra}
\end{figure*}
Figure \ref{FIG_spectra} shows exemplary $(\Psi,\Delta)$ spectra of the as-deposited coatings, acquired at an incidence angle $\theta$ = 60$^\circ$. At higher energy, above 5.7 eV, sample absorption is the cause of the observed degradation of the signal to noise ratio. The absorption could be related to the formation of colour centres, to the presence of roughness or to a poorly stoichiometric structure, where some aluminium particles might not be fully fluorinated \cite{Ode14}. We then used a pole in the UV region and a Tauc-Lorentz oscillator for the otical model of the thin film layer, which simultaneously fitted all the measured spectra with the same accuracy. A Bruggeman effective medium approximation (EMA) surface layer was added to the model to account for possible surface effects such as roughness.

\begin{figure}
 	\centering
 	\includegraphics[width=0.40\textwidth]{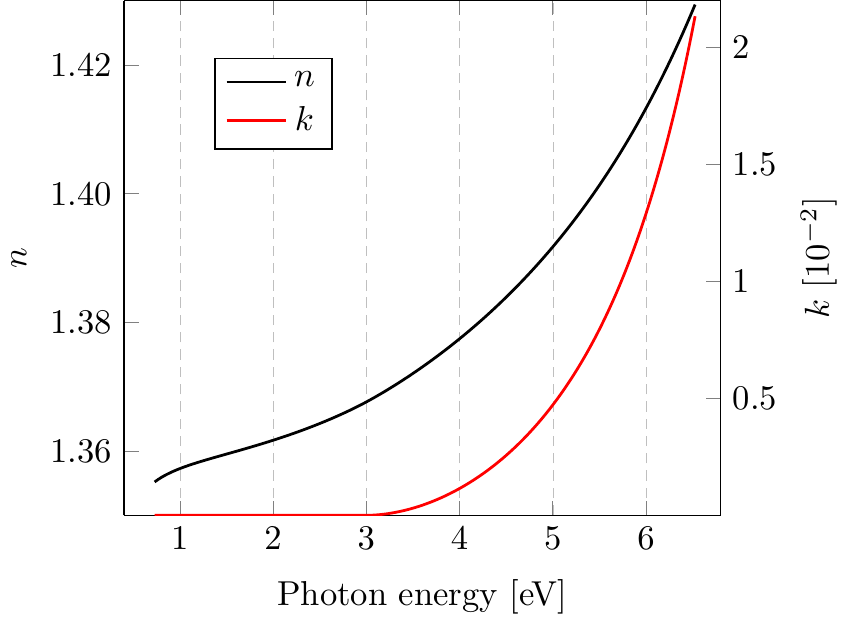}
 	\caption{Refractive index $n$ and extinction coefficient $k$ of as-deposited IBS AlF$_3$ thin films as a function of photon energy, derived from ellipsometric measurements. Relevant values for present and future GWDs are 0.80 and 1.17 eV, corresponding to a laser wavelength of 1550 and 1064 nm, respectively. For energy values smaller than $\sim$3.5 eV, the extinction is smaller than the sensitivity of the ellipsometers ($k < 10^{-3}$).}
 	\label{FIG_optConst}
\end{figure}

Figure \ref{FIG_optConst} shows the dispersion law and the extinction curve derived from our analysis of the as-deposited coatings, and Table~\ref{TABLEoptProp} lists our results (including density) against those we found in the literature concerning IBS AlF$_3$ thin films \cite{Kolbe93,Yoshida06,Ode14}. The Bruggeman layer was found to be almost $4 \%$ of the total coating thickness, indicating that the surface condition may play a role in optical loss at low wavelength. Values at $E$ = 1.17 eV and $E$ = 0.80 eV are particularly relevant, since those photon energies correspond to 1064 and 1550 nm, respectively, which are the operational laser wavelenghts of current and future GWDs \cite{aLIGO,AdVirgo,KAGRA,ET}. Refractive index values are $n = 1.358 \pm 0.005$ at 1064 nm and $n = 1.356 \pm 0.005$ at 1550 nm. For comparison, the refractive index at 1064 nm of the IBS silica coatings of present GWDs is $n = 1.47 \pm 0.01$ before annealing \cite{Granata20}.
\begin{figure}
 	\centering
 	\includegraphics[width=0.40\textwidth]{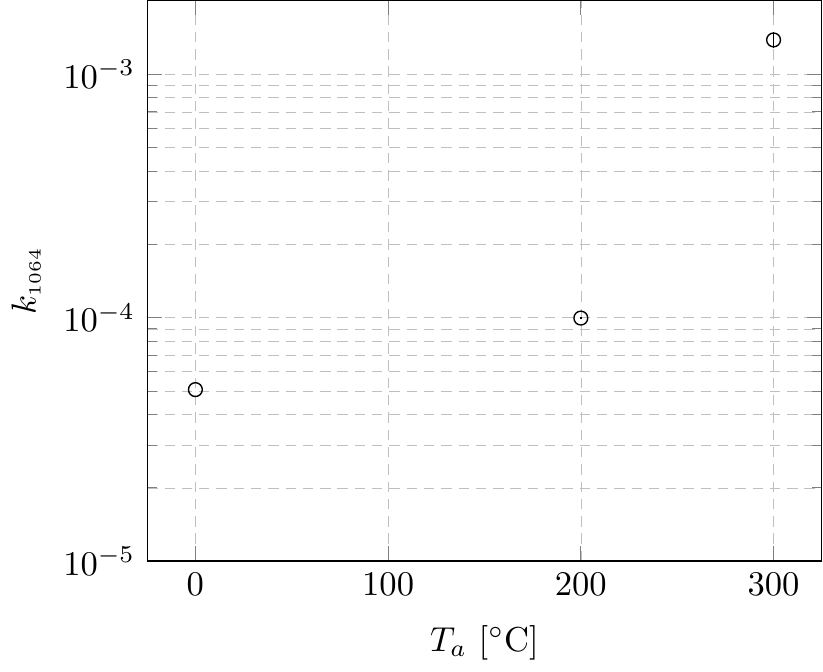}
 	\caption{Extinction coefficient $k_{\textrm{\tiny{1064}}}$ of IBS AlF$_3$ thin films as a function of the annealing temperature $T_a$, obtained from photo-thermal deflection measurements of optical absorption performed at 1064 nm ($T_a = 0$ $^{\circ}$C denotes as-deposited coatings).}
 	\label{FIG_absorptVSann}
\end{figure}
\begingroup
\squeezetable
\begin{table*}
	\caption{\label{TABLEoptProp} Refractive index $n$, extinction coefficient $k$ and density $\rho$ of as-deposited IBS AlF$_3$ thin films. $n$ and $k$ values presented in this work were measured in the 190-1700 nm wavelength range via spectroscopic ellipsometry. The $k$ value at 1064 nm was deduced from photo-thermal deflection measurements of optical absorption, by assuming negligible scatter loss. $^{(*)}$Extrapolations.}
	\begin{ruledtabular}
		\begin{tabular}{cccccccc}
			& & This work & Kolbe et al. \cite{Kolbe93} & Yoshida et al. \cite{Yoshida06} & Ode \cite{Ode14}\\ \cline{3-6}
			$n$ & \multirow{2}{*}{1550 nm} & 1.356 $\pm$ 0.005 & & \\
			$k$ & & $<10^{-3}$ & & \\ \cline{1-2}
			$n$ & \multirow{2}{*}{1064 nm} & 1.358 $\pm$ 0.005\\
			$k$ & & $(5.06 \pm 0.01)\times 10^{-5}$\\ \cline{1-2}
			$n$ & \multirow{2}{*}{633 nm} & 1.362 $\pm$ 0.005\\
			$k$ & & $<10^{-3}$\\ \cline{1-2}
			$n$ & \multirow{2}{*}{193 nm} & 1.4 $^{(*)}$ & 1.41 & 1.43 & 1.42\\
			$k$ & & $1.4\times 10^{-2}$ $^{(*)}$ & $5\times 10^{-4}$ & $1.3\times 10^{-4}$ & $8.3\times 10^{-6}$\\ \cline{1-2}
			$\rho$ [g/cm$^{3}$] & & 2.2 $\pm$ 0.4\\
		\end{tabular}
	\end{ruledtabular}
\end{table*}
\endgroup

Figure \ref{FIG_absorptVSann} shows the extinction coefficient $k$ obtained from the photo-thermal deflection measurements of optical absorption at 1064 nm, as a function of the annealing temperature $T_a$ using the measured thickness of 211 nm, as reported in Table~\ref{TABLEproperties}. 
We assumed that loss by light scatter was negligible.
We measured $k = 5.1 \times 10^{-5}$ before treatment, then $k$ increased after each annealing step up to $k = 1.4 \times 10^{-3}$ for $T_a = 300$ $^{\circ}$C.

\subsection{Mechanical properties}
It is known that post-deposition annealing treatment can improve mechanical properties of a material \cite{Granata20}, but it is important to establish both the ideal peak temperature $T_a$ and the duration $\Delta t_a$ at that temperature.

\subsubsection{Sample A}
Disk A underwent a series of annealing treatments of equal duration $\Delta t_a$ but of increasing plateau temperature $T_a$ (Table~\ref{TABLEsampleA}).
Heating and cooling ramps of 100 $^{\circ}$C every hour were used.
\begin{table} [htp]
	\caption{Annealing treatments applied to sample A.}
	\label{TABLEsampleA} 
	\begin{ruledtabular}
		\begin{tabular}{lccccccc}
			run & \#1 & \#2 & \#3 & \#4 & \#5 & \#6 & \#7\\ \hline
			$T_a$ [$^{\circ}$C] & 120 & 200 & 285 & 330 & 373 & 462 & 550\\
			$\Delta t_a$ [h] & 10 & 10 & 10 & 10 & 10 & 10 & 10\\
			\colrule
		\end{tabular}
	\end{ruledtabular}
\end{table}

\begin{figure*}[htp]
 	\centering
 	\includegraphics[width=0.78\textwidth]{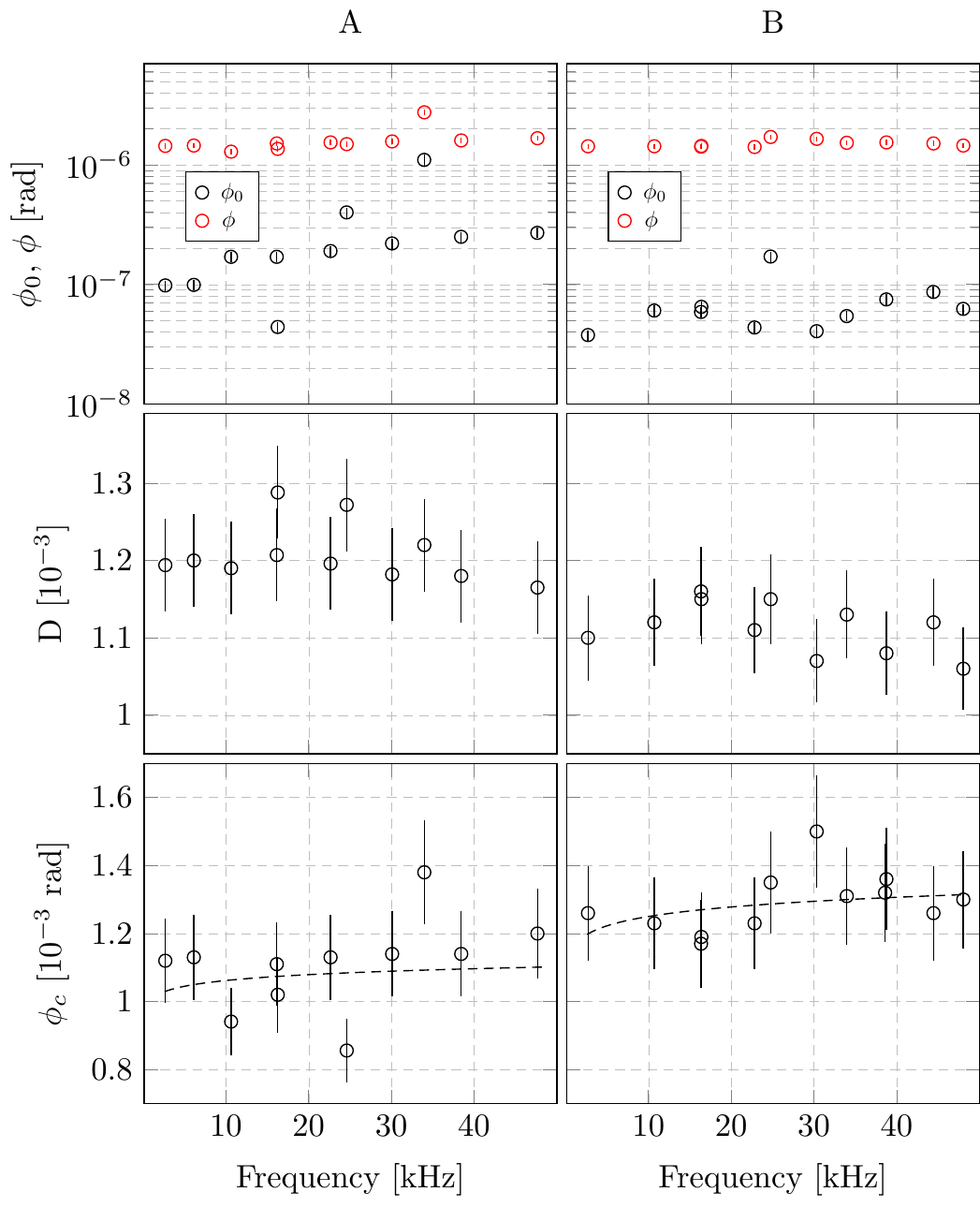}
 	\caption{
 		Characterization of mechanical losses, as function of frequency, of sample A (\textit{left column}) and sample B (\textit{right column}). 
 		\textit{Top row:} measured losses of the bare fused-silica substrate ($\varphi_0$) and coated substrate ($\varphi$).
 		\textit{Middle row:} frequency-dependent dilution factor
 		\textit{Bottom row:} coating loss angle ($\varphi_c$) of as-deposited IBS AlF$_3$. 
 		Most loss angle values of the two thin films are compatible within the error bars.
 		All measurements were performed by LMA before the first annealing treatment.
 		The best-fit power-law model of equation \ref{eq:power_law} are reported using a dashed line.
 	}
 	\label{fig:Dilution_Phi_A_B}
\end{figure*}

After each annealing we measured the loss angle $\varphi$ of the coated silica disk and applying equations~\ref{EQcoatLoss} and \ref{EQdilFact} we could measure the frequency-dependent dilution factor and extract the coating loss angle $\varphi_c$. The results achieved at LMA before the first annealing treatment are shown in \textit{Figure}~\ref{fig:Dilution_Phi_A_B}.
It is possible to describe the frequency-dependent behaviour of the coating loss angle fitting a power-law model to our data via least-squares linear regression \cite{Gilroy81,Travasso07,Cagnoli18}:
\begin{equation}
	\label{eq:power_law}
	\varphi_c (f) = a \left(\frac{f}{10~\text{kHz}}  \right)^b
\end{equation}

In Table~\ref{TABLE_parameters} are reported the mechanical properties of as-deposited IBS AlF$_3$ estimated using the procedure described in Section (II D) and the best-fit parameters (a,b) for each sample.
\begin{table} [htp]
	\caption{Measured Young's modulus $Y$ and Poisson ratio $\nu$ and best-fit parameters of the power-law model of equation~\ref{eq:power_law}}.
	\label{TABLE_parameters} 
	\begin{ruledtabular}
		\begin{tabular}{lcccc}
			 & Y [GPa] & $\nu$ & a [$10^{-3}$ rad] & b \\ \hline
			disk A & 78 $\pm$ 2 & 0.27 $\pm$ 0.02 & 1.06 $\pm$ 0.06 & 0.02 $\pm$ 0.06 \\
			disk B & 72 $\pm$ 1 & 0.22 $\pm$ 0.03 & 1.25 $\pm$ 0.04 & 0.03 $\pm$ 0.03 \\
			\colrule
		\end{tabular}
	\end{ruledtabular}
\end{table}

\begin{figure}[htp]
 	\centering
 	\includegraphics[width=0.45\textwidth]{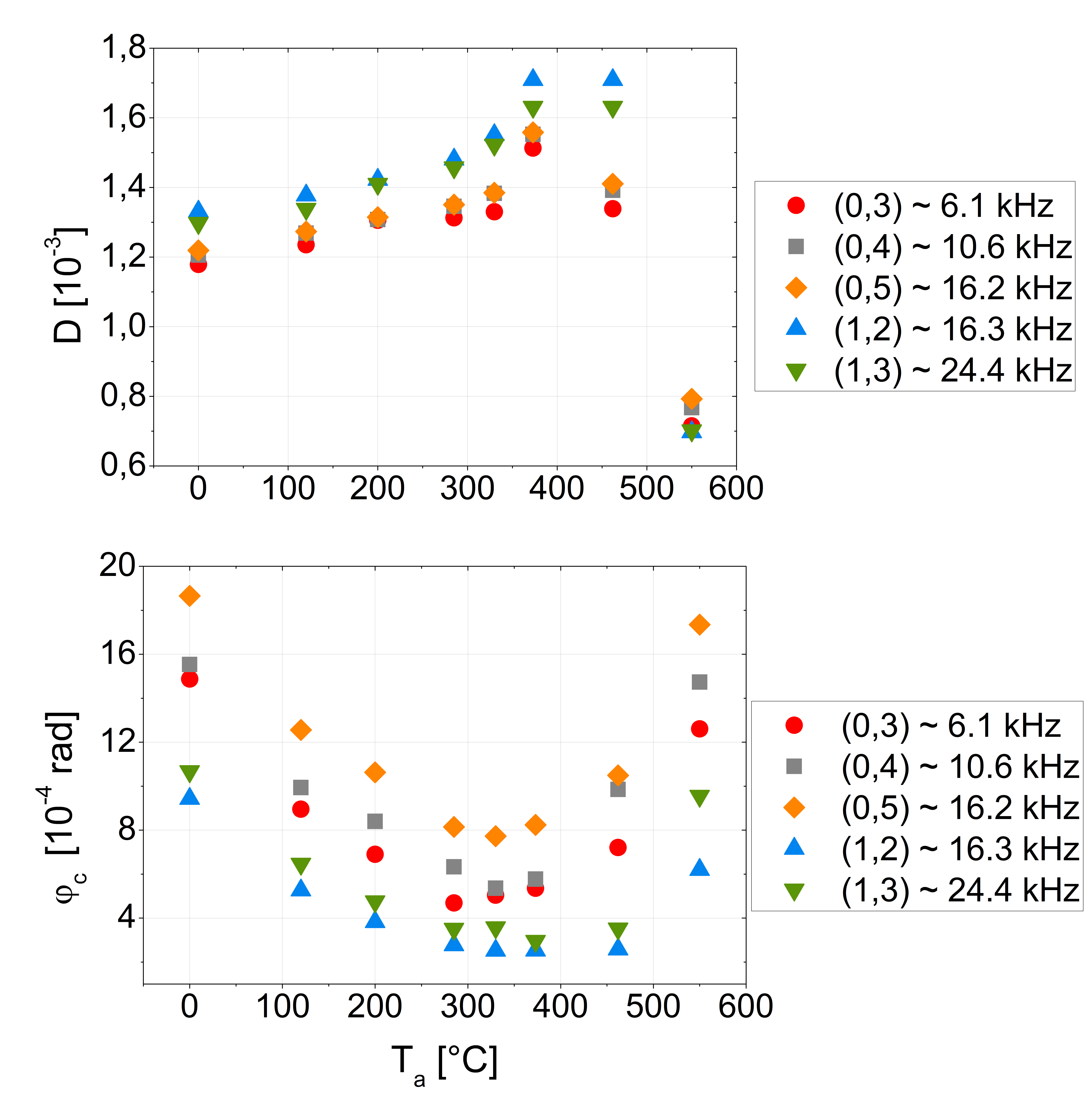}
 	\caption{Sample A: evolution of the dilution factor (top) and coating loss angle (bottom) with annealing temperature for some resonant modes. Errors in dilution factor measurements are $\sim 5 \%$ and $\sim 8 \%$ in loss angle measurements}
 	\label{fig:A_Dilution_Phi}
\end{figure}

\begin{figure}[htp]
 	\centering
 	\includegraphics[width=0.35\textwidth]{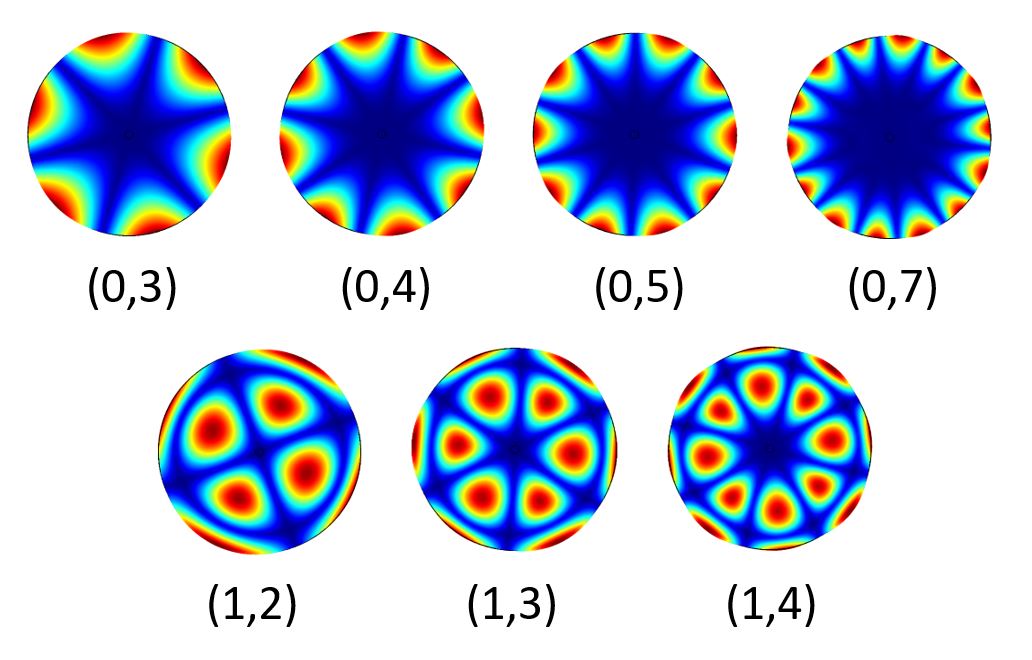}
 	\caption{Resonant modes identified with the $(r,a)$ notation. The colour scale represent the amplitude of oscillations from blue (no movement) to red (wider movement). Using a GeNS system it is possible to observe resonant modes with a nodal point in the centre.}
 	\label{fig:modes}
\end{figure}

In Figure~\ref{fig:A_Dilution_Phi} are reported the evolutions of dilution factor and coating loss angle with annealing temperature, measured at the UniUrb laboratory.
The first set of data ($T_a = 0$ $^{\circ}$C) represents the measured values before the first annealing. The $(r,a)$ notation denote different vibrational modes with $r$ radial and $a$ azimuthal nodes that are summarized in Figure~\ref{fig:modes}. 

From Figure~\ref{fig:A_Dilution_Phi} it is possible to see that the dilution factor increases until the treatment at 373 $^{\circ}$C and that the coating loss angle decreases after each annealing until the treatment at 373 $^{\circ}$C. The annealing temperature that minimizes the coating loss angle is between 285 $^{\circ}$C and 373 $^{\circ}$C.

The mass of the sample remains unchanged until the treatment at 373 $^{\circ}$C then starts to decrease of $0.3$~mg after the annealing at 462 $^{\circ}$C and $0.4$~mg after the last annealing at 550 $^{\circ}$C. So, after the last treatment, we have lost $0.7$~mg of coating out of $2$~mg. 

\begin{figure}[htp]
 	\centering
 	\includegraphics[width=0.26\textwidth]{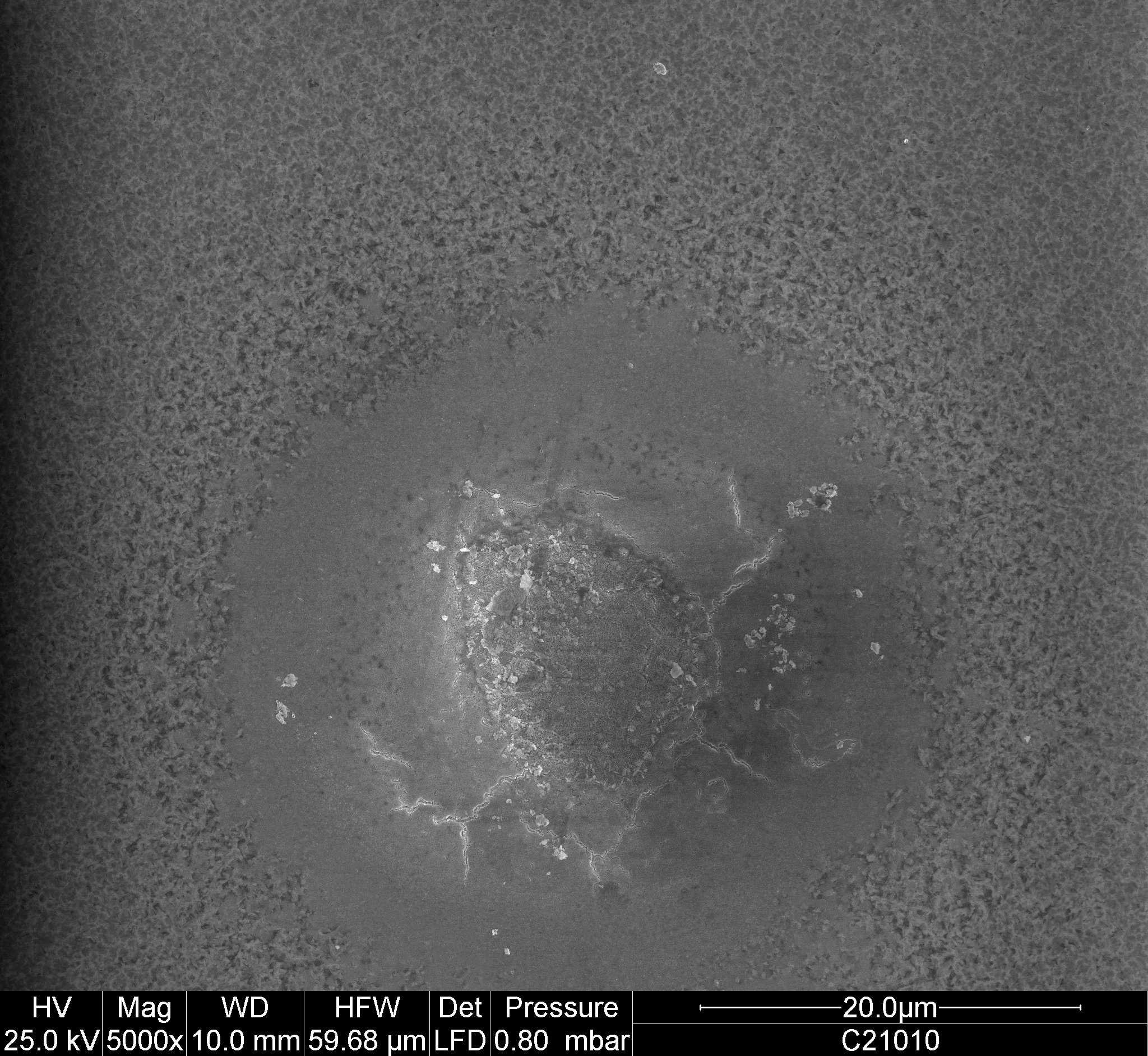}
 	\caption{Sample A: defect formation on the coating surface after the annealing at 330 $^{\circ}$C. Environmental scanning electron microscope: FEI quanta 200 FEG.  Magnification: 5000x}
 	\label{fig:A_bubbles}
\end{figure}

Moreover, using the optical microscope, we started to observe some round structures on the surface of the coating after the treatment at 330 $^{\circ}$C. An example, recorded using an environmental scanning electron microscope, is reported in Figure~\ref{fig:A_bubbles}.

\begin{figure}[htp]
 	\centering
 	\includegraphics[width=0.26\textwidth]{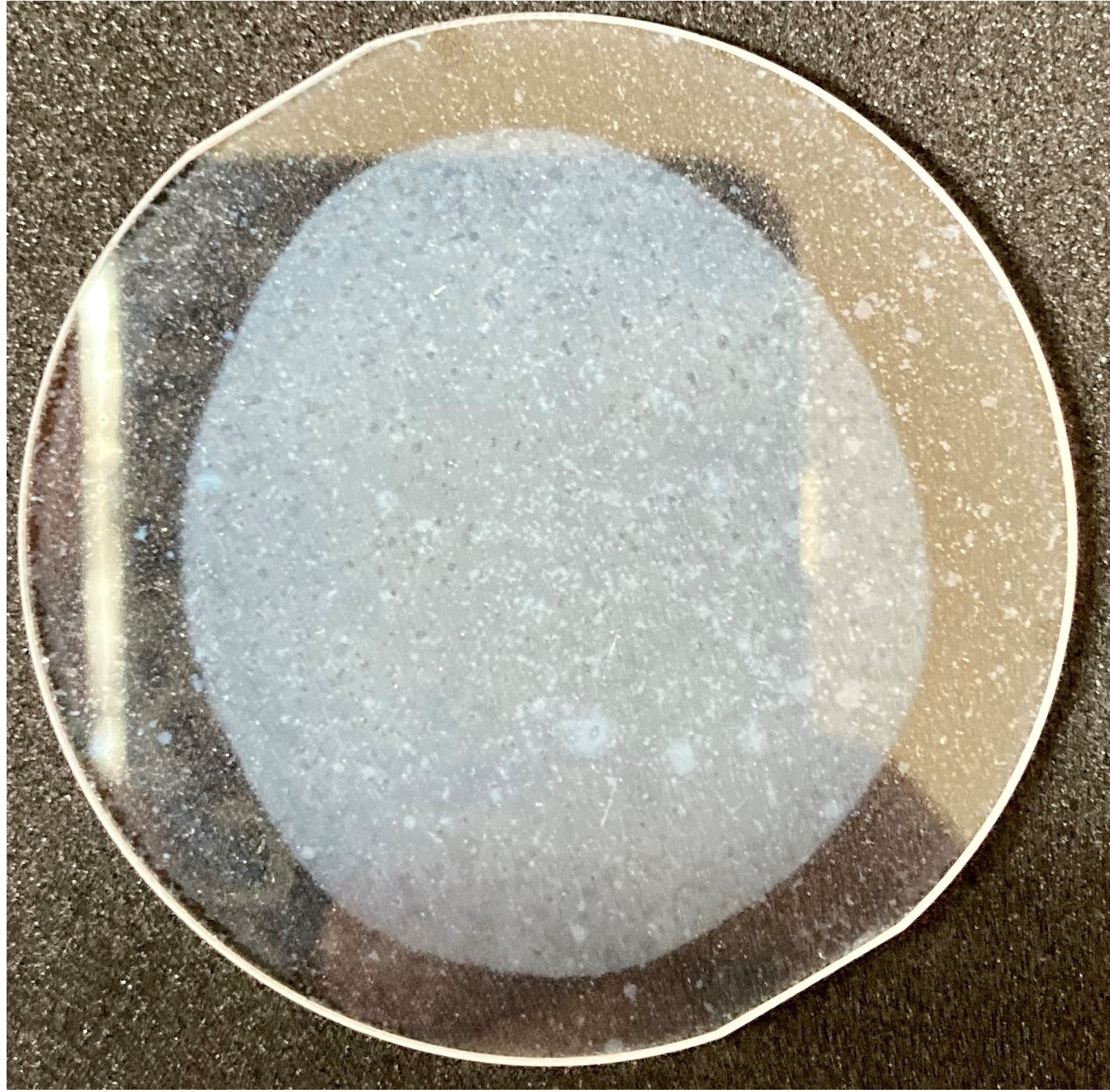}
 	\caption{Sample A: coating surface in contact with the quartz box after the annealing at 550 $^{\circ}$C.}
 	\label{fig:A_opaque}
\end{figure}

\begin{figure}[htp]
 	\centering
 	\subfloat[][\textit{}]
 	{\includegraphics[width=0.25\textwidth]{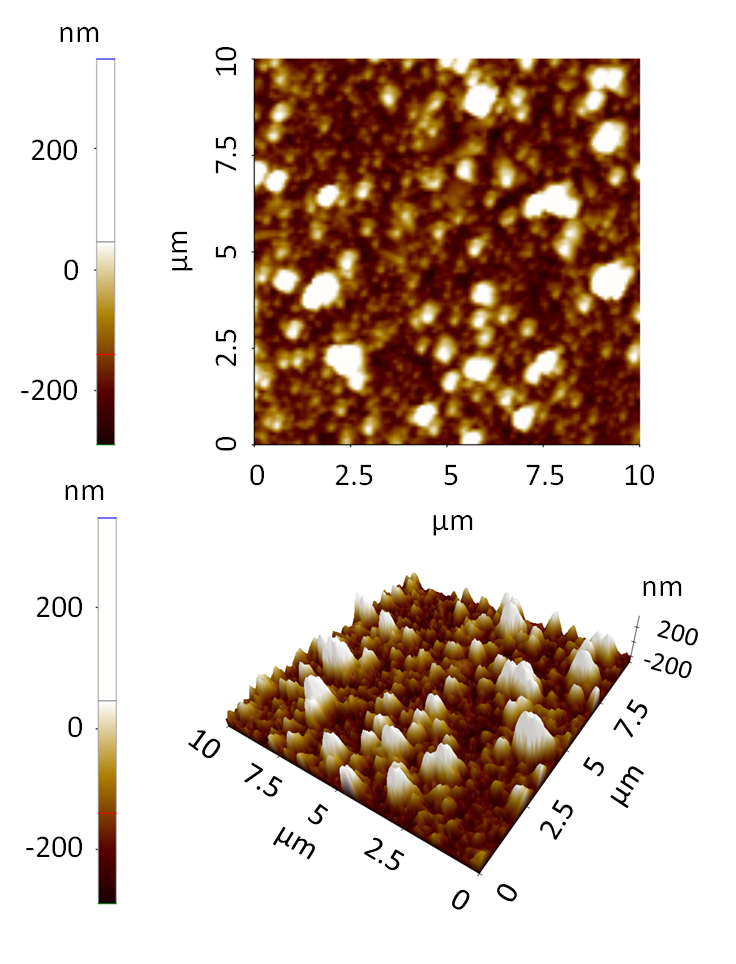}}
 	\subfloat[][\textit{}]
 	{\includegraphics[width=0.25\textwidth]{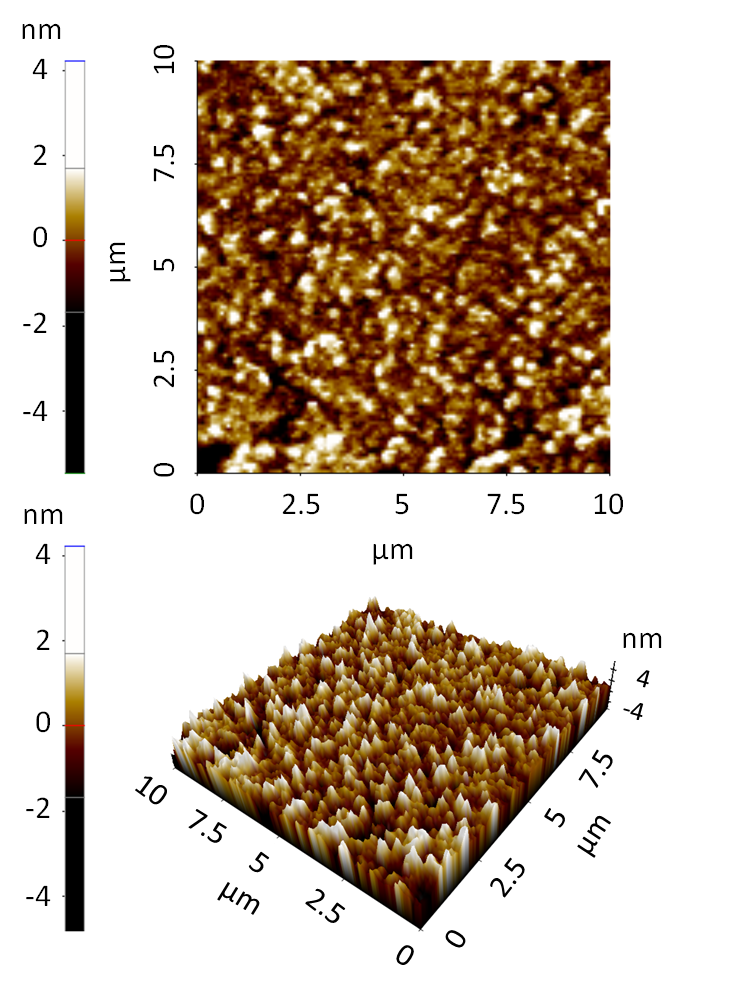}}
 	\caption{Morphology of the two sides of Sample A measured with AFM (XE100 Park System Inc): side~(a) was in contact with the quartz box during the annealing at 550 $^{\circ}$C and has a mean roughness of 67.21 nm; side~(b) has a mean roughness of 0.614 nm.}
 	\label{fig:A_AFM}
\end{figure}

After the treatment at 550 $^{\circ}$C the appearance of the coating changed: the side in contact with the quartz box used inside the furnace during the annealing looked opaque in the centre as shown in Figure~\ref{fig:A_opaque}. The morphology of the coating, measured with an atomic force microscope (AFM), appeared very different on the two sides of the sample:  the roughness of the side in contact with the quartz box reported in Figure~\ref{fig:A_AFM} (a) completely changed from its original value.

\begin{figure}[htp]
 	\centering
 	\subfloat[][\textit{}]
 	{\includegraphics[width=0.25\textwidth]{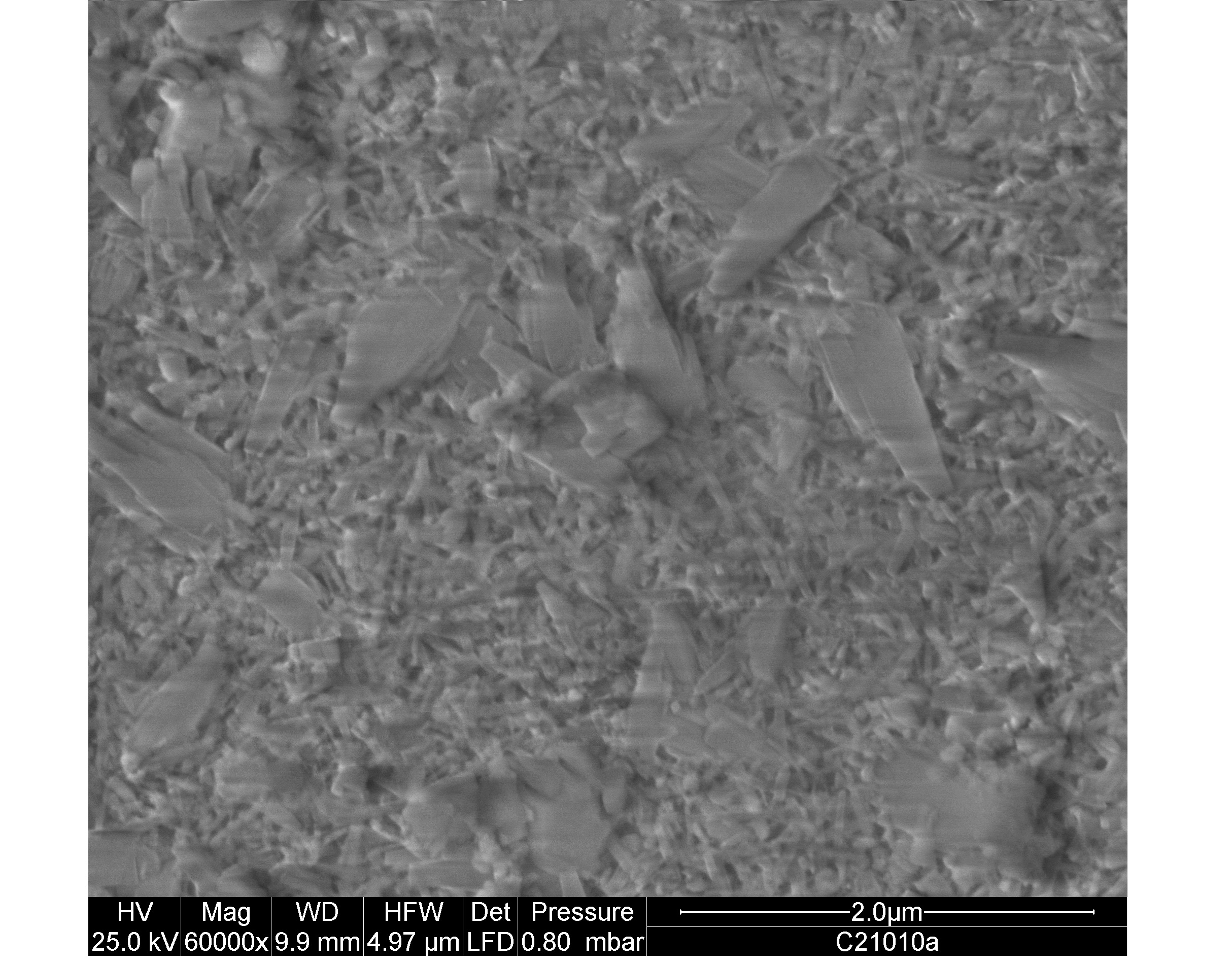}}
 	\subfloat[][\textit{}]
 	{\includegraphics[width=0.25\textwidth]{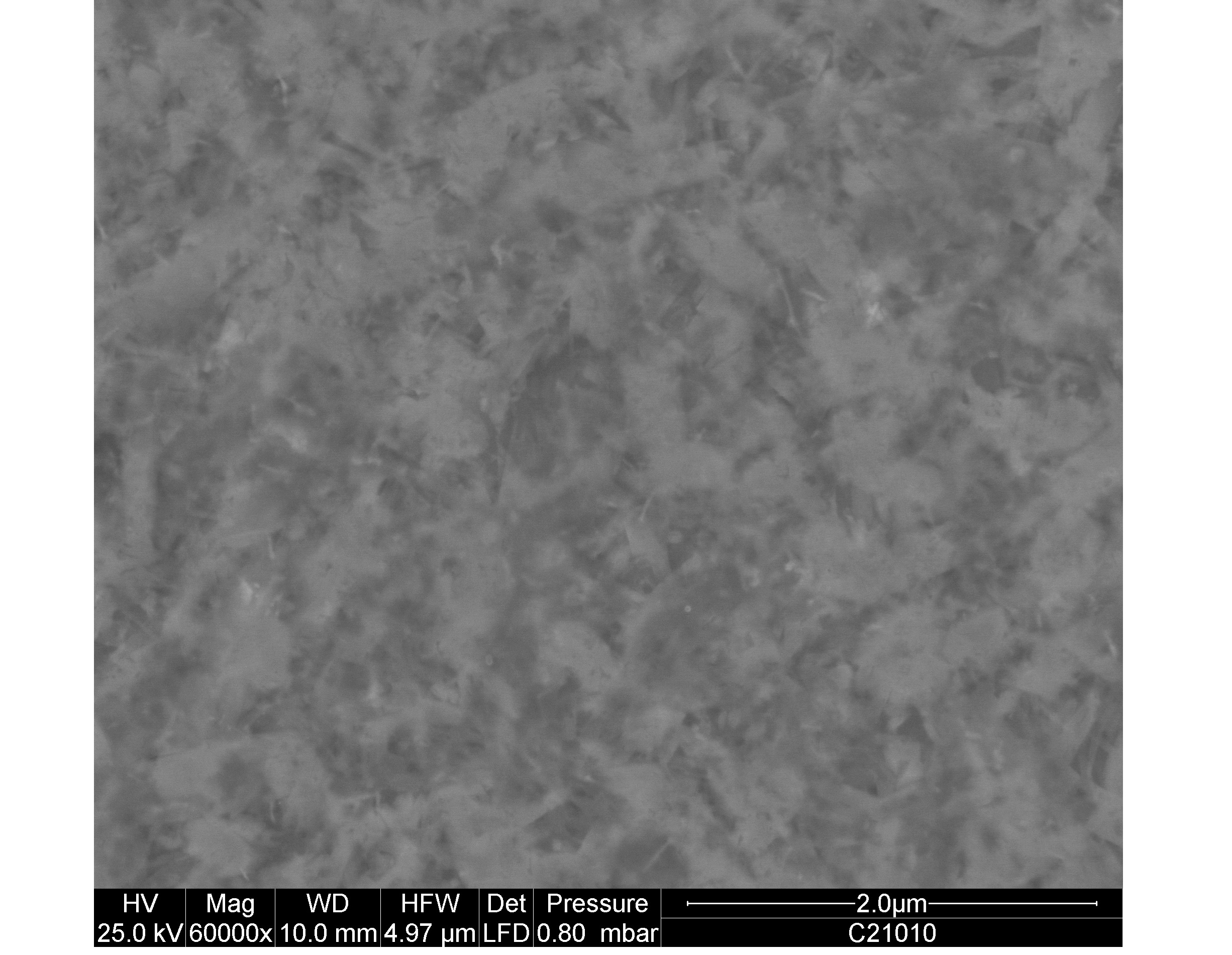}}
 	\caption{Crystals formation observed on the two sides of sample A using the enviromental scanning electron microscope (FEI quanta 200 FEG equipped with a spectrometer) Magnification: 60000x. Side~(a) was in contact with the quartz box during the annealing at 550 \textbf{$^{\circ}$C}.}
 	\label{fig:A_SEM}
 \end{figure}

Moreover, we observed the surface and the chemical composition of the sample with an enviromental scanning electron microscope (FEI quanta 200 FEG equipped with a spectrometer). In Figure~\ref{fig:A_SEM} it is possible to observe the formation of crystals on both sides of the sample. Crystals on side (a) are coarser than on side (b). In Table~\ref{TABLE_chemical_composition_A} we reported the chemical compositions of the two sides measured with the spectrometer. Stoichiometry of AlF$_3$ is not respected: there is a deficiency of fluoride on both sides of the sample. The presence of silicon and oxygen is due to the silica substrate under the coating. 
It seems that the fluoride reacted with the silicon contained in the quartz box during the thermal treatment at 550 $^{\circ}$C producing a volatile compound (SiF$_4$): this can explain the mass loss, the defects formation on the surface and the deficiency of fluoride.
\begin{table} [htp]
	\caption{Chemical composition of sample A measured with the spectrometer after the annealing at 550 $^{\circ}$C}
	\label{TABLE_chemical_composition_A} 
	\begin{ruledtabular}
		\begin{tabular}{lcc}
		Element & Atomic \% side(a) & Atomic \% side(b) \\ \hline
			O & 55.3 $\pm$ 0.6 & 52.6 $\pm$ 0.6 \\
			F & 4.4 $\pm$ 0.2 & 7.5 $\pm$ 0.4 \\
			Al & 24.8 $\pm$ 0.6 & 30.8 $\pm$ 0.5 \\
			Si & 15.5 $\pm$ 0.9 & 9.1 $\pm$ 0.7 \\
			\colrule
		\end{tabular}
	\end{ruledtabular}
\end{table}

\subsubsection{Sample B}
Disk B, nominally identical to disk A, underwent a series of annealing treatments of equal $T_a$ but of increasing $\Delta t_a$ (Table~\ref{TABLEsampleB}). We have indicated as "cumulative duration" the total amount of time at the plateau temperature.
\begin{table} [htp]
	\caption{Annealing treatments applied to sample B.}
	\label{TABLEsampleB}
	\begin{ruledtabular}
		\begin{tabular}{lccc}
			run & \#1 & \#2 & \#3\\ \hline
			$T_a$ [$^{\circ}$C]& 285 & 285 & 285\\
			$\Delta t_a$ [h] & 10 & 20 & 30\\
			cumulative duration [h] & 10 & 30 & 60\\
			\colrule
		\end{tabular}
	\end{ruledtabular}
\end{table}

\begin{figure}[htp]
	\centering
	\includegraphics[width=0.45\textwidth]{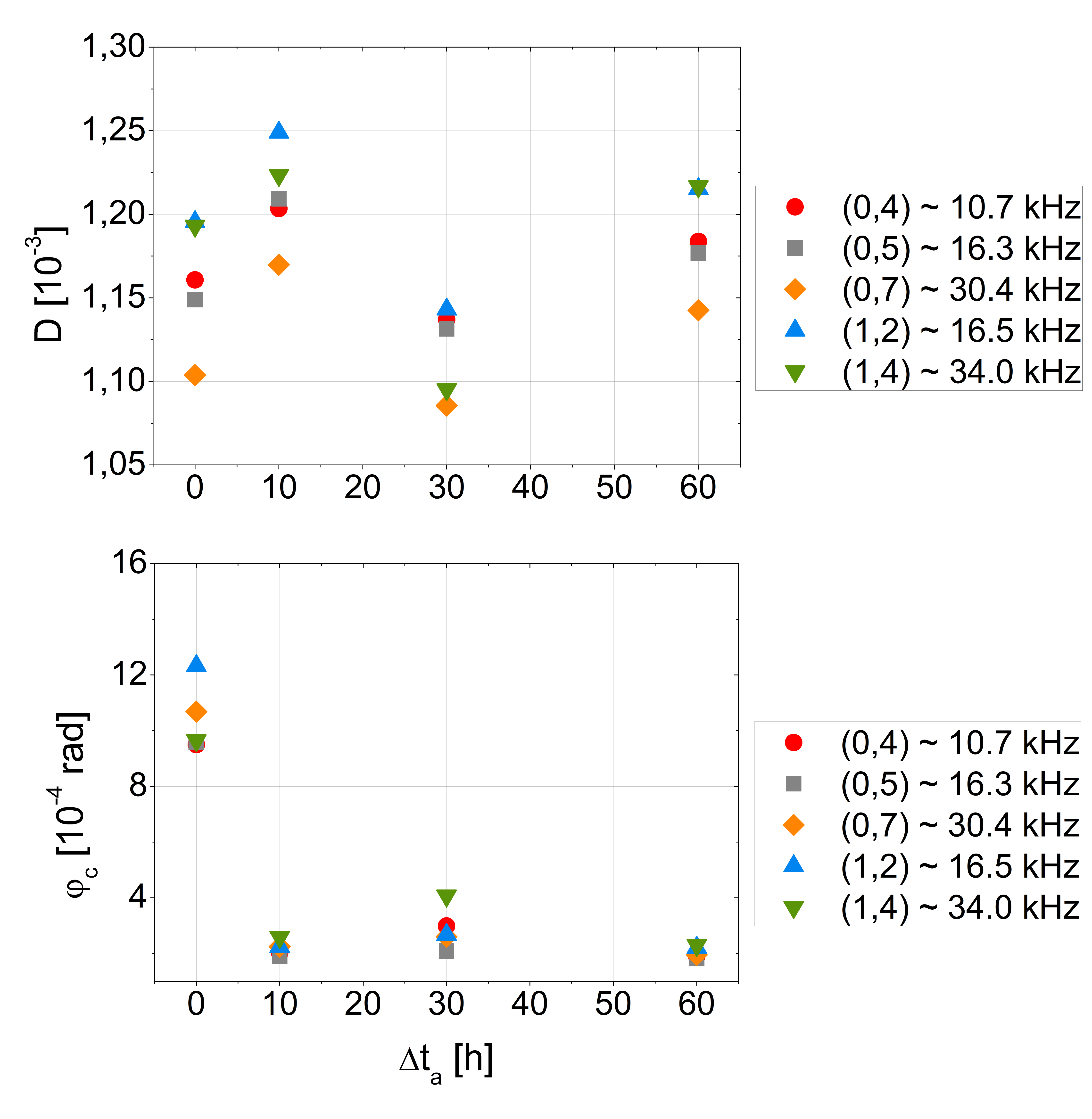}
 	\caption{Sample B: evolution of the dilution factor (top) and coating loss angle (bottom) with annealing temperature for some resonant modes. Errors in dilution factor measurements are $\sim 5 \%$ and $\sim 8 \%$ in loss angle measurements}
 	\label{fig:B_Dilution_Phi}
\end{figure}

From the results on Sample A, we have chosen 285 $^{\circ}$C as the best annealing temperature, because it minimizes the coating loss angle and seems not to produce defects on its surface.
After each annealing we measured the coated disk and applying equations~\ref{EQcoatLoss} and \ref{EQdilFact} we could measure the frequency-dependent dilution factor and extract the coating loss angle $\varphi_c$. The results are reported in Figure~\ref{fig:B_Dilution_Phi}. The first set of data ($\Delta t_a = 0$ $^{\circ}$C) represents the measured values (at UniUrb laboartory) before the first annealing.
The coating loss angle decreased after the first treatment, but no more improvements were observed after subsequent annealings.

The mass of the sample remained unchanged after all thermal treatments. No defects or crystals were observed using the optical microscope and the environmental scanning electron microscope.
	\begin{table} [htp]
		\caption{Chemical composition of sample B measured with the spectrometer after the last annealing at 285 $^{\circ}$C}
		\label{TABLE_chemical_composition_B} 
		\begin{ruledtabular}
			\begin{tabular}{lcc}
				Element & Atomic \% side(a) & Atomic \% side(b) \\ \hline
				O & 19.1 $\pm$ 0.4 & 15.2 $\pm$ 0.5 \\
				F & 50.8 $\pm$ 0.4 & 56.1 $\pm$ 0.5 \\
				Al & 28.7 $\pm$ 0.3 & 27.5 $\pm$ 0.4 \\
				Si & 1.4 $\pm$ 0.3 & 1.2 $\pm$ 0.2 \\
				\colrule
			\end{tabular}
		\end{ruledtabular}
	\end{table} 

 In Table~\ref{TABLE_chemical_composition_B} we reported the chemical compositions of the two sides measured with the spectrometer. In this case, there is more fluoride than aluminium, however the stoichiometry is not respected.

\section{Conclusions}
In the framework of a research activity devoted to find low-noise coating materials for present and future gravitational-wave detectors, we characterized the optical and mechanical properties of a set of IBS AlF$_3$ thin films. We chose fluoride coatings because of their low refractive index $n_{\textrm{\tiny{L}}}$, with the aim of minimizing the overall high-reflection coating thickness $t_c$ in Eq.(\ref{eqn.S}). As a reminder, $t_c$ is a monotonically decreasing function of the refractive index contrast $C = n_{\textrm{\tiny{H}}}/n_{\textrm{\tiny{L}}}$. Furthermore, because of their potentially low mechanical loss at low temperature \cite{Schwarz11}, fluorides could be a valid option for use in cryogenic detectors.

Indeed, the IBS AlF$_3$ thin films featured a lower refractive index at 1064 nm than that of IBS silica layers of current detectors \cite{Granata20}. However, their optical absorption and ambient-temperature loss angle turned out to be considerably higher.

Concerning the coating loss angle, the best annealing temperature is found to be 285 $^{\circ}$C (from the analysis of \textit{Sample A}) and a duration of 10 hours at the plateau temperature is sufficient to reach a minimum in the coating loss angle $\varphi_c \simeq 2.2 \cdot 10^{-4}$ rad achieved in sample B.
Higher annealing temperatures could produce a chemical reaction between the fluoride of the coating and the silicon contained in the substrate, producing defects on the sample surface.
However, the loss angle of AlF$_3$ coatings at room temperature is still too high compared to silica coatings, whose value is $\varphi_c \simeq 2.3 \cdot 10^{-5}$ rad \cite{Granata20}.
Further tests will be carried out in the near future to explore the mechanical behaviour of AlF$_3$ coatings at low temperatures, where the silica loss angle increases and should be replaced by another low refractive index material.

From a point of view of optical properties, the optical absorption increases after each annealing and is at least 3 orders of magnitude higher than what is tolerable for GWDs applications \cite{Granata20}. In the future, efforts will be made to optimise the quality of the coating material in order to decrease the coating optical absorption and loss angle. Lower optical and mechanical losses could possibly be achieved by changing the coating growth conditions \cite{Granata20}, as well as by improving the stoichiometry and reducing the amount of impurities.

\section{Acknowledgments}
This work has been promoted by the Laboratoire des Mat\'{e}riaux Avanc\'{e}s and partially supported by the Virgo Coating Research and Development (VCR\&D) Collaboration. The authors would like to thank M. Gauch, F. Carstens and H. Ehlers of the Laser Zentrum Hannover for the production of the AlF$_3$ thin films and for the fruitful discussions. In the online document repositories of the LIGO and the Virgo Scientific Collaborations, this work has been assigned document numbers LIGO-P2100478 and VIR-1385B-21, respectively.

\end{document}